# Field tuned critical fluctuations in YFe$_2$Al$_{10}$: Evidence from magnetization, $^{27}$Al (NMR, NQR) investigations

P. Khuntia[1,*], A. Strydom[2], L. S. Wu[3,4], M. C. Aronson[3,4], and F. Steglich[1], and M. Baenitz[1]

[1]Max Planck Institute for Chemical Physics of Solids, 01187 Dresden, Germany

[2]Physics Department, University of Johannesburg, P.O. Box 524, Auckland Park 2006, South Africa

[3]Condensed Matter Physics and Materials Science Department, Brookhaven National Laboratory, New York 11973, USA

[4]Department of Physics and Astronomy, Stony Brook University, Stony Brook, New York 11794, USA

We report magnetization, specific heat, and NMR investigations on YFe$_2$Al$_{10}$ over a wide range in temperature and magnetic field and zero field (NQR) measurements. Magnetic susceptibility, specific heat and spin-lattice relaxation rate divided by $T$ (1/T$_1$T) follow weak power law ($\sim T^{-0.4}$) temperature dependence, which is a signature of critical fluctuations of Fe moments. The value of the Sommerfeld-Wilson ratio and linear relation between $1/T_1T$ and $\chi$ suggest the existence of ferromagnetic correlations in this system. No magnetic ordering down to 50 mK in $C_p(T)$ and the unusual $T$ and $H$ scaling of the bulk and NMR data are associated with a magnetic instability which drives the system to quantum criticality. The magnetic properties of the system are tuned by field wherein ferromagnetic fluctuations are suppressed and a crossover from quantum critical to FL behavior is observed with increasing magnetic field.

PACS: 71.27.+a, 74.40.Kb, 76.60.-k, 76.60.Es, 71.10.Hf

Critical fluctuations (CF) close to the absolute zero in temperature in materials as diverse as heavy-fermion systems [1-4], cuprates [6,7], and iron pnictides [7] intimate the proximity to a quantum critical point (QCP). Whether the quantum critical point is underpinned by a magnetic phase transition or not depends on the location of the material on the generic phase diagram that connects a paramagnetic ground state with antiferromagnetic (AFM) order through an intervening QCP. The dimensionality of the material has recently been demonstrated as an insightful new materials classification approach in the physics of quantum criticality (QC) [8]. Controlling a magnetic phase transition by means of a tuning parameter such as chemical composition, magnetic field ($H$) or pressure has proven to be a profitable route to QC [4,9]. Non-Fermi-liquid (NFL) behavior of thermal properties is a hallmark of QC. The $T$ and $H$ scaling of magnetic systems are an insightful approach to unravel detail such as the universality of quantum phase transitions (QPT). For instance, unusual scaling in the resistivity $\rho(T) \sim T^n$ (n<2), specific heat $C_p(T)/T \sim [-\ln T$ or $1-\alpha T^{-n}(n<1)]$ and magnetic susceptibility $\chi(T) \sim T^n$ (n<1) at low $T$ are associated with QPT [1-4]. A QPT typically causes a pile-up of entropy towards $T \to 0$. The NFL behavior in the local probe NMR leads to unusual power laws in spin-lattice relaxation rates (SLR) with $1/T_1T \sim T^n$ (n<2) in the $T \to 0$ limit in contrast to simple local moment metals wherein n=0 is expected. From a theoretical point of departure, NFL behavior displaying these $T$ scaling relations are native to instabilities in ferromagnetic (FM) and AFM systems, and spin fluctuations (SF) may also display similar NFL-like divergences. The concept of QC in 4$f$- and 3$d$-electron spin-density-wave systems in which an appropriate tuning parameter dissolves AFM order into a paramagnetic FL is by now well established [1-4, 10,11]. For instance, in $YbRh_2Si_2$, small fields suppress the AFM order and $1/T_1T \sim T^{-0.5}$ is observed, which is due to a delicate interplay between FM and AFM SF of local $Yb^{3+}$ moments in the vicinity of a QCP [12,13]. The significance of SF and their field dependence in itinerant FM like MnSi [14], $Y(Co_{1-x}Al_x)_2$ [15], $ZrZn_2$ [16], $Ni_3AlC_x$ [17], $Sr_{1-x}Ca_xRuO_3$ [18], $Sc_3In$ [19] and LaCoPO [20,21] have been explained by NMR in the framework of the self consistent renormalization (SCR) theory [22]. The notion of QC when the order parameter is FM on the other hand is controversial and currently a subject of debate. Recently, weak FM systems like $UGe_2$ [23], MnSi [24], $ZrZn_2$ [25], and $NbFe_2$ [26,27] have been showing exciting features and it is observed that the magnetic field and/or pressure act as tuning parameters. Furthermore, unusual and weak power laws in $\chi(T)$ and $C_p(T)/T$ due to atomic disorder in a few itinerant and local moment systems are interpreted by quantum Griffith phase model [28-37]. The scaling behaviors of especially $T$ and $H$ offer a systematic approach towards establishing universality classes. In QC systems however our current understanding of the energy scaling that drives cooperative behavior offers very little commonality. The QC region of phase space appears to offer increasing layers of detail which complicates a consistent treatment of the QC region among different materials. The notion of universality still appears to be a distant target of our understanding of QC.

Ternary orthorhombic 4$f$-aluminides of $RT_2Al_{10}$ (R:Y,Yb,Ce; T: Fe, Ru,Os) type are currently attracting much interest because of a number of exotic properties such as unconventional structural and magnetic ordering in $CeRu_2Al_{10}$, metamagnetic transition as well as Kondo-insulating behavior in $CeFe_2Al_{10}$ [38-41]. In $YFe_2Al_{10}$, the most recently investigated member of this series, there is mounting evidence that this compound is situated very close to a QCP of FM nature. The proximity to FM ordering is conveyed especially through an Arrott plot presentation of magnetization [42,43], but $YFe_2Al_{10}$ presents no unambiguous evidence of magnetic order above 50 mK.

In this communication, we present our investigations on magnetic susceptibility, specific heat, $^{27}$Al NMR and NQR data in polycrystalline (PC) and single crystalline (SC) $YFe_2Al_{10}$. We have performed temperature and magnetic field dependent measurements in order to connect local magnetic dynamics through NMR and NQR with bulk thermal entropy and the consequences of FM correlations [42,43]. Bulk data indicate that $YFe_2Al_{10}$ is in the vicinity of a magnetic instability [42,43]. NMR investigations in the $T$ range 1.8 $\leq T \leq$ 300 K and in the H range 0$\leq \mu_0 H$ $\leq$7.27 T on the PC sample have been carried out to provide better insights concerning the low energy spin excitations, and to contribute a deeper understanding of the effect of field on the correlated spin dynamics. Furthermore, we succeeded to perform $^{27}$Al NQR measurements, which reflect the intrinsic magnetism of the system.

Details pertaining to the synthesis of PC and SC samples of $YFe_2Al_{10}$ are described elsewhere [42,43]. The careful thermal and microprobe analysis inferred high quality and stoichiometric composition of the system without significant atomic disorder [44]. We found no evidence for site exchange and only one Fe site is present in the host lattice [44]. This poses a route towards understanding the QC in this system without complicating factors of atomic disorder whether accidental or whether induced as a consequence of doping when used to suppress FM order that is observed in other putative QC systems. The magnetic susceptibility $\chi(T)$ is measured in different $H$ in the $T$ range 1.8≤$T$≤300 K in PC samples using a Quantum Design (QD), MPMS SQUID magnetometer.

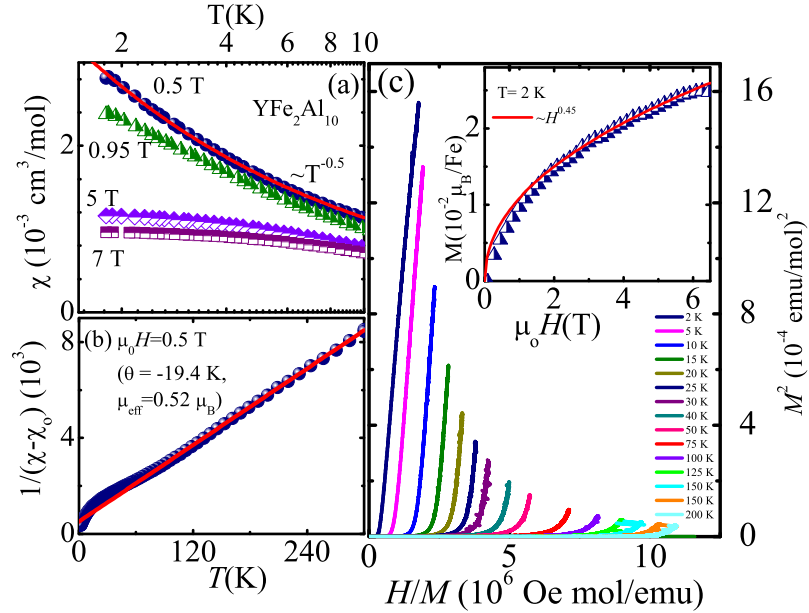

Fig. 1. (Color online). (a) Temperature dependence of $\chi(T)$ in different magnetic fields. The solid line displays the $T^{-0.5}$ behavior. The inset shows the T dependence of $1/\chi(T)$ with C-W fit (c) Arrott plot of the magnetization. The inset shows $M$ vs. $H$ at 2 K with a fit to $H^{0.45}$.

As shown in Fig.1a, the low field $\chi(T)$ increases upon lowering $T$ which could be due to exchange enhanced $q=0$ excitations. The ac and dc magnetic susceptibility in small magnetic fields follow $T^{-0.5}$ behavior below 10 K, indicating the presence of spin correlations and critical spin fluctuations. However, this power law in $\chi(T)$ is different that observed in the SC, which can be associated with some sort of powder averaging in the PC sample [43]. The high field $\chi(T)$ tends to saturate below 5 K, which is due to suppression of fluctuation of Fe moments. Above 50 K, $\chi(T)$ obeys a Curie-Weiss behavior (see Fig. 1b) and the effective moment ($\mu_{eff}$= 0.52 $\mu_B$/f.u) is rather small. The low value of saturation moment ($\mu_s \sim 0.02$ $\mu_B$/f.u) leads to $\mu_{eff} / \mu_s \approx 26$, which classifies YFe$_2$Al$_{10}$ being a weak itinerant FM in the Rhodes-Wolfrath plot [45,46]. The magnetization isotherm at 2 K displays $H^{0.45}$ (as shown in Fig. 1c inset), which implying critical SF. $M$ vs. $H$ follows a mean field type $M^2$ vs. $H/M$ scaling (Fig. 1c), suggesting $T\rightarrow 0$ FM ordering. The nature of the Arrott plot (see Fig. 1c) is attributed to the zero-point CF of Fe moment close to QCP, which are suppressed with magnetic field.

Specific heat on the PC samples was measured in zero field down to 50 mK, using the dillution refrigerator and in various fields down to 0.35 K in the $^3$He option of QD PPMS. The specific heat coefficient $C_p(T)/T$ at low $T$ is enhanced (see Fig.4a) and follow a $T^{-0.35}$ behavior, indicating quasi-2D ferromagnetic CF of Fe moments at $H$=0. It is noteworthy that in YFe$_2$Al$_{10}$, Fe atoms are stacked in 2D planes perpendicular to the *b*-direction of the orthorhombic crystal structure and the CF's reside largely within the *ac*-plane [43]. The low temperature $C_p(T)/T$ at 7 T is corrected for the nuclear Schottky contributions arising mainly from $^{27}$Al [47]. The Sommerfeld coefficient, $\gamma = C_p(T)/T$ is enhanced ($\sim$14.5 mJ/mol Fe K²) compared to $\gamma \sim$3 mJ/mol Ru K² in the isostructural YRu$_2$Al$_{10}$ [43] indicating quasi-particle mass enhancement and suggests that the system is driven from a critical regime to a FL state with field. The value of the Sommerfeld-Wilson ratio R$_w$==$\pi^2 k_B^2/\mu_0\mu_{eff}(\chi/\gamma)$ ($\approx$7 at 2 K) is relatively large implying the presence of FM correlations. These findings could be ascribed to the effect of SF in this weak itinerant FM wherein SF modulates the thermo-dynamical observables greatly [22]. The low $T$ increase of $C_p(T)/T$ in YFe$_2$Al$_{10}$ could be associated with the virtual scattering of electrons on the Fermi surface via $q$=0 components of SF and critical SF's implying an enhancement of the effective mass of electrons [4,9]. The exchange enhancement and mass enhancement are attributed to FM instability in the close proximity of a QCP [22]. The specific heat in a metal is related to the density of states, which seems to be modified around the QCP leading to unusual scaling laws in $T$ and $H$ which is substantiated in NMR and NQR data as discussed in the following sections [22].

The presence of five inequivalent Al sites yields very complex NMR spectra (see Fig. 2). Each Al site has different quadrupole couplings which makes the powder spectra rather broad. No appreciable shift is observed in NMR spectra down to 1.8 K. The low frequency spectra broaden inhomogeneously which originates from the broad distributions of *H*-independent quadrupole interactions admixed with the *H*-dependent anisotropic Zeeman interactions. The line broadening at low *T* is associated with fluctuations of the Fe moments, and consequent strong magnetic correlations, which are in agreement with macroscopic data [22,48].

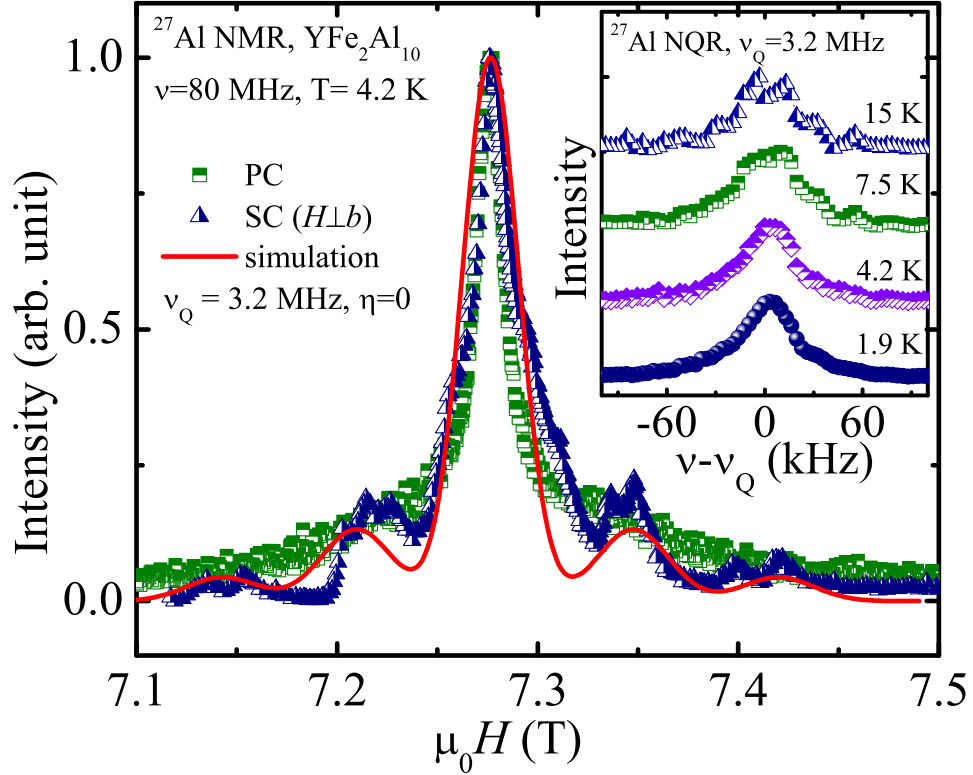

Fig. 2. (Color online). (a) Field sweep NMR PC and SC spectra (*H*⊥*b*) with simulation at 80 MHz. The inset shows the NQR spectra ( ±3/2↔ ±5/2) on PC sample at different temperatures.

The quadrupole Hamiltonian is given as:

$$H_Q = \frac{h\nu_Q}{6}\left[(3I_z^2 - I^2) + \frac{\eta}{2}(I_x^2 - I_y^2)\right]$$

Where $\eta=(V_{xx}-V_{yy})/V_{zz}$ is the asymmetry parameter ($\eta=0$), $\nu_Q=3eQV_{zz}/20h$ is the NQR frequency, $V_{zz}$ is the EFG tensor, which arises from the local charge distribution. The system presented here has five Al sites resulting in ten NQR lines since each Al site has two NQR lines viz., $\pm 1/2 \leftrightarrow \pm 3/2$; $\pm 3/2 \leftrightarrow \pm 5/2$ transitions. Fig. 2 shows the 4.2 K field sweep NMR spectra at 80 MHz for PC together with SC ($H \perp b$) samples. The SC NMR spectra exhibit more pronounced (first order) quadrupolar transitions with lines corresponding to the aforementioned pair of transitions. The central transition is rather broad for a SC but we interpret this as being due to strong correlations that are prevalent. To simplify the present spectra with five different Al sites (and five different quadrupole coupling constants, $\nu_q$) we simulated NMR spectra with single Al site. This simulation yields $\nu_q=1.6$ MHz which corresponds to the fraction of Al site having the highest quadrupolar interactions. For these Al sites, NQR lines (for $\eta=0$) are expected at $\nu_{Q1}=\nu_q$ ($\pm 3/2 \leftrightarrow \pm 1/2$) and $\nu_{Q2}=2\nu_q$ ($\pm 5/2 \leftrightarrow \pm 3/2$). We therefore performed NQR at 3.2 MHz and we indeed observed a relatively narrow NQR line (see Fig. 2. inset) which is in good agreement with our field sweep NMR results. Above 10 K, the NQR line develops singularities which are symptomatic of multiple quadrupolar transitions. Moreover, the well resolved NQR lines enable us to probe the low temperature magnetic fluctuations via SLR measurements in $H=0$. Because of the relatively weak $T$ and $H$ dependence of the line width and the absence of a sizable shift of the spectra and the SLR measurements could reliably be performed over a wide $T$ and $H$ range of our investigation and the analysis of which is discussed in the following sections. Almost no shift in NMR and NQR spectra could be attributed to the symmetrical position of Fe with respect to Al nuclei and very weak hyperfine fields resulting in cancellation of transverse hyperfine fields via $^{27}$Al.

$^{27}$Al SLR measurements have been performed by exciting the central transition (which is not affected by the quadrupole interactions) following saturation recovery method with suitable rf pulses in different $H$. The recovery of longitudinal magnetization $M(t)$, at a time delay t after the saturation pulse in all $T$ and $H$ ranges could consistently be fitted with a single component valid for $I=5/2$ nuclei: $1-M(t)/M(\infty)=0.0291e^{-t/T_1}+0.178e^{-6t/T_1}+0.794e^{-15t/T_1}$, where $M(\infty)$ is the equilibrium magnetization [49]. The fit of $M(t)$ to the single component suggests the uniform distribution of Fe moments in the host lattice. $1/T_1$ decreases with decreasing $T$ and shows linear behavior with $T$ in high magnetic fields, which is attributed to the metallic nature of the system

due to the effect of non interacting conduction electrons. This leads to $T$-independent $1/T_1T$ behavior in high $H$, and the system is in the FL state. A cross-over from Korringa to QC behavior in $1/T_1T$ is observed in weak $H$ and $1/T_1T$ increases monotonically and follows $T^{-0.4}$ behavior below 100 K as shown in Fig. 3a evidencing the persistence of CF of weak itinerant moments in weak $H$ [15,20]. The observed power law divergence in SLR of YFe$_2$Al$_{10}$ suggests a fragile interplay between FM ($q=0$) and AFM ($q\neq0$) components of SF's [12,13,22].

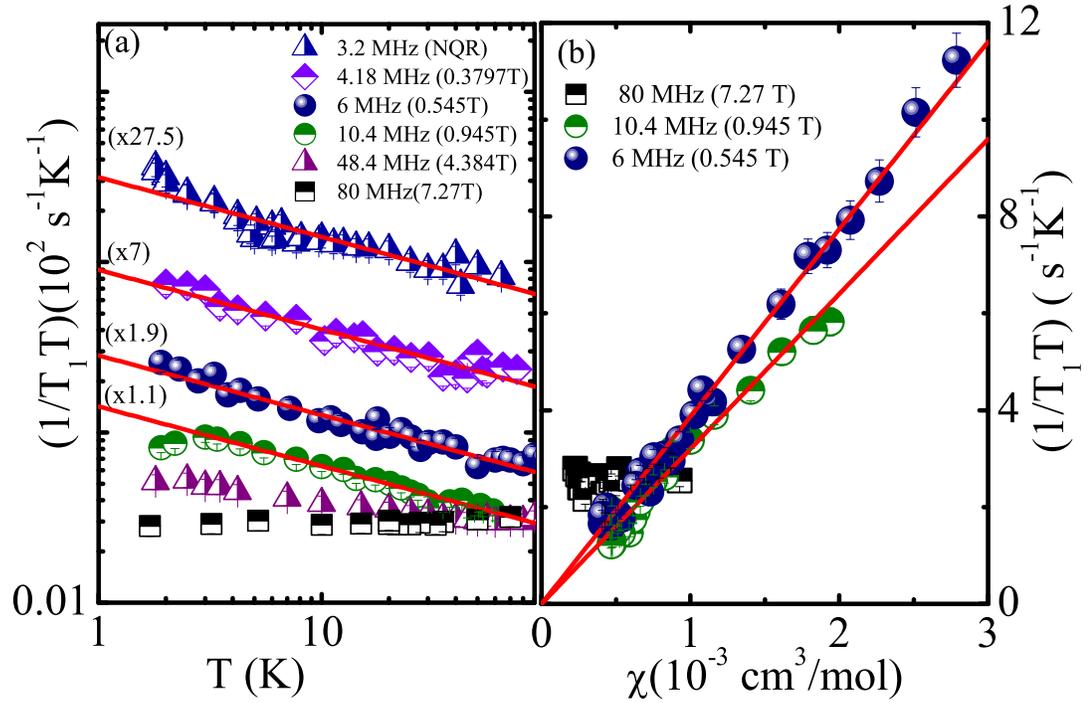

Fig. 3.(Color online).(a) $T$ dependence of $1/T_1T$ in various magnetic fields and the solid line is a fit to $T^{-0.4}$. ($1/T_1T$ values are scaled for better clarity) (b) The linear relation between $1/T_1T$ and $\chi$ with $T$ as an implicit parameter.

SLR ($1/T_1T$) in principle probes the $q$-averaged low energy spin excitations and it can be expressed as the wave-vector $q$ summation of the imaginary part of the dynamical spin susceptibility $\chi''(q,\omega_n)$ [50]:

$$\frac{1}{T_1T} = \frac{2(^{27}\gamma_n)^2}{N_A^2} k_B \sum_q |A_{hf}|^2 \frac{\chi''(q,\omega_n)}{\omega_n}$$

where $\chi''(q,\omega_n)$ is the dynamic electron spin susceptibility and $A_{hf}(q)$ is the hyperfine form factor.

The CF of itinerant moments lead to the observation of weak power laws in $1/T_1T$ vs. $T$, which is a consequence of the magnetic instability that is caused by hybridization between nearly filled *d*-electron shells and conduction electrons. The SLR is linear with $\chi(T)$ in weak magnetic fields (Fig. 3b), which evidence the dominant role of FM SF [18, 22, 51, 53]. It is worth to mention that $1/T_1T \sim \chi(T)$ behavior represents the fact that $\chi(T)$ is the uniform spin susceptibility $\chi(q=0, \omega=0)$ and accounts for the dominant FM component of the static spin susceptibility [22,50]. $1/T_1T$ reflects the dynamic SF which is not related to bulk effects, and the suppression of $1/T_1T$ with $H$ might be attributed to the fact that the relaxation mechanism is governed by the intrinsic magnetism of the system. The FM correlations develop between Fe moments, and the fluctuation of local hyperfine fields of electronic states with respect to $^{27}$Al nuclei offers a channel for the relaxation mechanism at the FM wave vector $q=0$. Modes of SF's are localized in reciprocal space with small amplitude in YFe$_2$Al$_{10}$ leading to unusual behavior in SLR, which infer that the system is in the vicinity of a QCP [13,22]. NMR investigations on the SC are limited due to the poor signal to noise ratio in the available tiny SC and skin depth problems. Nonetheless, the values of $T_1$ in SC are found to be close to those in the PC and infer the absence of strong anisotropy in $T_1$ for both crystallographic directions $H \perp b$ and $H \| b$.

NQR SLR are measured for the $\pm 3/2 \leftrightarrow \pm 5/2$ transitions (with $\eta \sim 0$) following the saturation recovery method for the $M(t)$. $T_1$ is determined by fitting the recovery of $M(t)$: $1 - M(t)/M(\infty) = 0.427 e^{-3t/T_1} + 0.573 e^{-10t/T_1}$, where $M(\infty)$ is the equilibrium magnetization [49,54]. Zero field (NQR) $1/T_1T$ is associated with intrinsic CF, the $T$ dependence of which follow the same power law as in the case of low field $1/T_1T$ data as shown in Fig. 3a.

As mentioned previously, the close proximity of this system to a QCP could lead to unusual behavior of bulk and NMR data that are connected with magnetic interactions and collective excitations. Interestingly, the observation of same power law in $1/T_1T$ and $C_p(T)/T$ is not unusual as these type of features also observed in a few systems [10,30,31,55]. We rule out the possibility of Griffith phase in YFe$_2$Al$_{10}$, because of the absence of significant atomic disorder [44].

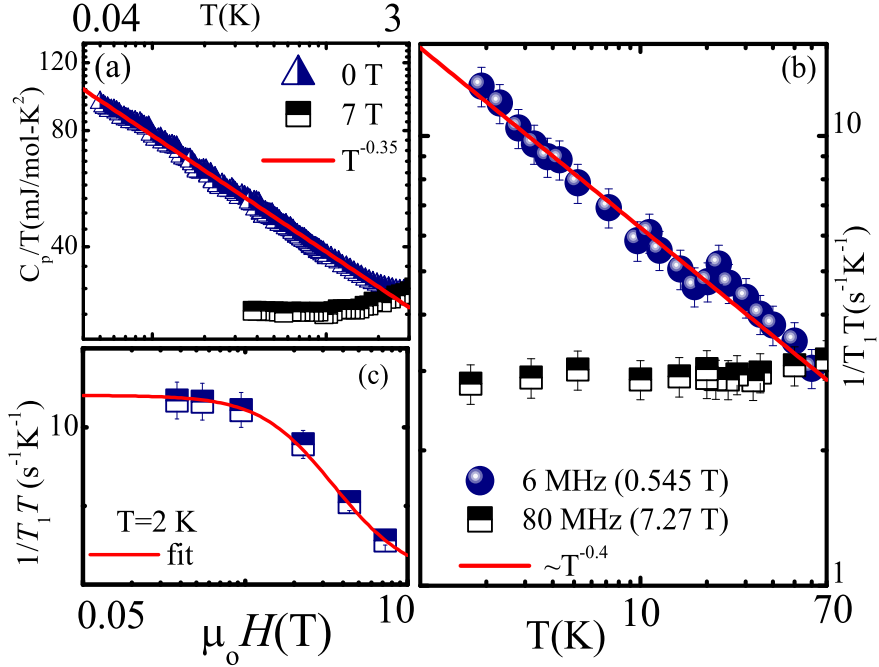

Fig. 4. (Color online). Temperature dependence of $C_p(T)/T$ and the solid line a fit to $T^{-0.35}$ (b) $1/T_1T$ vs. $T$ at 0.545 and 7.27 T (c) Field dependence of $1/T_1$ at 2 K with a fit as discussed in the text

It may be noted that the observed CF in $\chi$, $C_p(T)/T$ and $1/T_1T$ are suppressed with magnetic field suggesting the suppression of low lying spin excitations in the SF spectra. A magnetic field of 7 T is strong enough to dampen the fluctuations and diminishing the spin entropy towards a FL phase in YFe$_2$Al$_{10}$ [56-58]. The $H$-dependence of the SLR rate at 2 K follows a dynamic scaling $1/T_1(H)=1/T_1(0)[1+(\mu_B H/k_B T)^2]^\xi$, $\xi=0.74$ (see Fig. 4c). This scaling behavior is ascribed to the dominant role of the field in tuning the CF of spins in inducing the system to a FL phase [59-62]. The magnetic field accounts for the renormalization of collective excitations and suppresses the correlation effects in YFe$_2$Al$_{10}$, leading to suppression of spin fluctuations [59-62].

To summarize, the divergences observed in $\chi(T)$ and $C_p(T)/T$ of YFe$_2$Al$_{10}$ are consistently attributed to the quantum CF of Fe moments and support $T\to 0$ phase transition. Absence of magnetic ordering down to 50 mK and low $T$ and $H$ divergences in $\chi$, $C_p(T)/T$ and $1/T_1T$ cannot be interpreted in terms of the standard theory predicted for disordered or NFL model. Bulk and NMR data are consistent with the persistence of FM correlations and infer the dominant role of $q=0$ excitations and critical spin fluctuations in the spin excitation spectra in tiny magnetic fields. The observed static and dynamic magnetic properties are associated with a nearly continuous magnetic phase transition and the system seems to be a *d*-electron analogue of 4*f*-electron heavy fermions. The SF's which are of FM nature in $YFe_2Al_{10}$ are suppressed and a cross over from a QC regime to a FL state is observed, and the magnetic field is concluded to play the significant role in driving such a phase transition. Further investigations concerning the effect of chemical pressure and SLR measurements at low temperature are highly essential. Neutron scattering experiments are required to address the nature of spin fluctuation spectra along various *q* vectors.

We acknowledge insightful discussions with H. Yasuoka, M. Garst and M. Brando. We thank C. Klausnitzer for technical support concerning specific heat measurements and U. Burkhardt for microprobe and thermal analysis of the sample. We thank the DFG for financial support (OE-511/1-1). Work at Stony Brook University was carried out under NSF grant DMR-0907457.

——————————————————

*pkhuntia@gmail.com